\documentstyle[prl,preprint,aps,epsfig]{revtex}
\tightenlines
\begin{document}

\newcommand{\adag}{a^{\dag}}
\newcommand{\atil}{\tilde{a}}
\def\frp#1{${#1\over2}^+$}
\def\frm#1{${#1\over2}^-$}
\def\g{\noindent}

\def\mev{\hbox{\ MeV}}
\def\kev{\hbox{\ keV}}
\def\lambdabar{{\mathchar'26\mkern-9mu\lambda}}
\def\lambdabarrr{{^-\mkern-12mu\lambda}}
\def\lsim{\mathrel{\rlap{
\lower4pt\hbox{\hskip-3pt$\sim$}}
    \raise1pt\hbox{$<$}}}     
\def\gsim{\mathrel{\rlap{
\lower4pt\hbox{\hskip-3pt$\sim$}}
    \raise1pt\hbox{$>$}}}     

\draft
\title{First-forbidden mirror $\beta -$decays in $A=17$ mass region}

\author{N. Michel\dag, J. Oko{\l}owicz\dag\S, F. Nowacki\ddag, 
~and M. P{\l}oszajczak\dag\copyright}
\address{\dag\ Grand Acc\'{e}l\'{e}rateur National d'Ions Lourds (GANIL),
CEA/DSM -- CNRS/IN2P3, BP 5027, F-14076 Caen Cedex 05, France}
\address{\S\ Institute of Nuclear Physics, Radzikowskiego 152,
PL - 31342 Krakow, Poland}
\address{\ddag\ Laboratoire de Physique Th\'{e}orique  Strasbourg (EP 106),
3-5 rue de l'Universite, F-67084 Strasbourg Cedex, France }
\address{\copyright Institute for Nuclear Theory, University of Washington, Box
351550, Seattle, WA 98195, USA}


\maketitle

\begin{abstract}
\parbox{14cm}{\rm The first-forbidden $\beta -$decay of $^{17}$Ne into the 
'halo' state $J^{\pi}=1/2_1^+$ of $^{17}$F presents one of the largest measured 
asymmetries for mirror $\beta -$decay feeding bound final states. 
This asymmetry is
studied in the framework of the Shell Model Embedded in the Continuum (SMEC). 
The spatial extent of single particle orbits is
constrained by the proton capture cross-section $^{16}$O$(p,\gamma )^{17}$F
calculated in SMEC. This allows to estimate mirror symmetry breaking in 
$^{17}$F/$^{17}$O and
$^{17}$Ne/$^{17}$N nuclei.} 

\end{abstract}
\bigskip

\vfill
\newpage

\section{Introduction}
A realistic account of the low-lying states properties in exotic nuclei
requires taking into account the coupling between discrete and continuum states
which is responsible for unusual spatial features of these nuclei. This aspect is
particularly important in the studies near drip line where the amount of
experimental information is strongly reduced and one has to use both
structure and reaction data to understand basic properties 
of these nuclei. Within the Shell Model Embedded in the 
Continuum (SMEC) approach
\cite{bnop1,bnop2}, one may obtain the unified description of the 
divergent characteristics such as the spectra (energies of states, transition
probabilities, proton/neutron emission widths, $\beta -$decays, etc.) and
the reactions involving one-nucleon in the continuum 
(proton/neutron capture processes,
Coulomb dissociation reactions, elastic/inelastic proton/neutron reactions,
etc.). This provides a stringent
test of the effective interactions in the SMEC calculations and
permits to asses the mutual complementarity of reaction and structure data
for the understanding of these nuclei. In this context, it can be 
interesting to
compare the first-forbidden $\beta -$decay transition of $^{17}$Ne in the 
ground state (g.s.) $J^{\pi}=1/2_1^-$ to the weakly bound, first excited state 
$J^{\pi}=1/2_1^+$ in $^{17}$F, with the corresponding mirror 
decay of $^{17}$N into a well bound excited state of $^{17}$O. In these decays
an abnormal asymmetry of mirror decay rates has been observed 
by Borge et al. \cite{borge} and later confirmed by Ozawa et al. \cite{ozawa}. 
Borge et al. \cite{borge} explained this effect by the large
asymmetry of radial sizes of $s_{1/2}$ s.p. orbits involved in bound states of
$^{17}$F/$^{17}$O and $^{17}$Ne/$^{17}$N. Different explanation has been provided by Millener
\cite{millener} who attributes the bulk of the asymmetry to charge dependent
effects which lead to different $1s_{1/2}$ occupancy for the initial states
, {\it i.e.} to different amplitudes of $\pi(0p_{1/2}^21s_{1/2}^2)\nu(0p_{1/2}^{1})$
and $\nu(0p_{1/2}^21s_{1/2}^2)\pi(0p_{1/2}^{1})$ components in
the g.s. of $^{17}$Ne and $^{17}$N, respectively.

In the framework of SMEC, we shall discuss the constraints 
from proton capture cross-section data on the radial 
wave function involved  in the description of the first-forbidden 
$\beta -$decay into bound final states. 

In the SMEC formalism, which derives from the continuum shell model formalism
\cite{weid,bartz3}, the subspaces of (quasi-) bound (the $Q$ subspace) and
scattering (the
$P$ subspace) states are separated using the projection operator technique. 
$P$ subspace contains asymptotic channels, which are made of
$(N-1)$-particle localized states 
and one nucleon in the scattering state, whereas $Q$ subspace contains 
many-body localized states which are
build up by the bound single-particle (s.p.) wave functions
and by the s.p.\ resonance wave functions. 
The wave functions in $Q$ and $P$ are then properly renormalized
in order to ensure the orthogonality of wave functions in both subspaces.

In the first
step, we calculate the (quasi-) bound many-body states in $Q$ subspace. For
that we solve the multiconfigurational Shell Model (SM) problem :
$H_{QQ}{\Phi}_i = E_i{\Phi}_i ~ \ ,$ where $H_{QQ}$ is given by the SM
effective interaction which is responsible 
for the {\it internal mixing} of many-body
configurations. The quality of the SMEC
description depends crucially on the realistic account of  
configuration mixing for the coexisting low-lying structures and hence on the
quality of the SM effective interactions and the SM space
considered. In the description of $A=17$ 
nuclei, it is important to take into account
the dynamics of $^{16}$O core and to include 2p-2h and 4p-4h excitation from
$p-$shell to $sd-$ shell. Zuker-Buck-McGrory (ZBM) effective SM interaction 
in the basis of $0p_{1/2}$, $1s_{1/2}$ and $0d_{5/2}$ 
orbitals \cite{zbm,zbm1}, allows to take into account this dynamics.
The valence space ($0p_{1/2}$, $1s_{1/2}$, $0d_{5/2}$) has the advantage
to be practically non spurious and most of states at the 
$p-sd$ interface are well described through the
configuration mixing of these three orbitals. 
Results of this paper have been obtained with the ZBM interaction.

\section{Details of the model}
To generate the radial s.p.\ wave functions in $Q$ subspace
and the scattering wave functions in $P$ subspace, as a first guess,
we use the average potential of Woods-Saxon (WS) type
with the spin-orbit : $ V_{SO} {\lambdabar}_{\pi}^2 (2{\bf l}\cdot{\bf s})
r^{-1}d{f}(r)/dr$, and Coulomb parts included.  
${\lambdabar}_{\pi}^2 = 2\,$fm$^2$ is the pion Compton wavelength and
$f(r)$ is the spherically symmetrical WS formfactor.
The Coulomb potential $V_C$ 
is calculated for the uniformly charged sphere of radius $R_0$ (see Table 1).
This 'first guess' potential $U(r)$, is then modified by the
residual interaction. We shall return to this problem below.

 For the continuum part, we solve the coupled channel equations :
\begin{eqnarray}
\label{cchan}
(E - H_{PP}){\xi}_{E}^{(+)} \equiv
\sum_{c^{'}}^{}(E - H_{cc^{'}}) {\xi}_E^{c^{'}(+)} = 0 ~ \ ,
\end{eqnarray}
where index $c$ denotes different channels and $H_{PP} \equiv PHP$.
The superscript $(+)$ means that boundary
conditions  for incoming wave in the channel $c$ and
outgoing scattering waves in all channels are used.
The channel states are defined by coupling of one
nucleon in the scattering continuum to the many-body SM state in
$(N - 1)$-nucleus. For the coupling between bound and scattering
states around $^{16}\mbox{O}$, we use 
the density dependent interaction (DDSM1) \cite{bnop3,bnop4}.
This interaction provides {\it
external mixing} of SM configurations via the virtual excitations of
particles to the continuum states. The channel - channel coupling potential :
$H_{cc^{'}} = (T + U ){\delta}_{cc^{'}} + {\upsilon }_{cc^{'}}^{J} ~ \ ,$
contains the kinetic-energy operator $T$ and
the channel-channel coupling ${\upsilon }_{cc^{'}}^{J}$ generated by the
residual interaction. At a first
step, the potential for channel $c$ consists of the initial WS potential 
$U(r)$, and of the diagonal
part of coupling potential ${\upsilon }_{cc}^{J}$
which depends on both the s.p.\ orbit
${\phi}_{l,j}$ and the considered many-body state $J$. Hence,
the initial potential is modified by the coupling potential and in the next
step the s.p.\ wave functions ${\phi}_{l,j}$ defining $Q$ subspace are generated
by the modified potential, what in
turn modifies the diagonal part of the residual force, {\it etc.}
In other words, the procedure of solving coupled channel equations
(\ref{cchan}) is
accompanied by the self-consistent iterative procedure which for each
total $J$ yields the $J-$dependent self-consistent potential :
$U^{(sc)}(r) = U(r)+{\upsilon }_{cc}^{J(sc)}(r) ~ \ ,$
and consistent with it the new renormalized formfactor of the
coupling force. $U^{(sc)}(r)$ differs significantly from the initial potential,
especially in the interior of the potential \cite{bnop2,bnop3}. For weakly
bound many-body states, strong modification of the surface features of the initial
potential $U(r)$ has been found as well \cite{bnop2,bnop3}. 
Parameters of the first guess potential $U(r)$ are such that $U^{(sc)}(r)$
reproduces energies of experimental s.p.\ states,
whenever their identification is possible. For certain $J$, 
the s.p. wave functions are  not modified by the above iterative procedure. For
example in $J^{\pi}=1/2^+$ states of $^{17}$F and $^{17}$O, the
couplings to the g.s. $0^+$ of $^{16}$O modify only
$1s_{1/2}$ s.p. wave function. In this case, 
the radial wave functions of all other s.p. states
are generated by an auxiliary average potential $U^{(aux)}(r)$ 
of the WS type with spin-orbit and Coulomb parts included. 
We shall return to this problem below.

 The third system of equations in SMEC consists of the inhomogeneous
coupled channel equations:
\begin{eqnarray}
\label{ethird}
(E^{(+)} - H_{PP}){\omega}_{i}^{(+)} = H_{PQ}{\Phi}_i \equiv w_i ~ \ .
\end{eqnarray}
The source term $w_i$ depends on the
structure of $N$ - particle SM wave function ${\Phi}_i$. 
The radial formfactor of the source depends on s.p.\ wave functions of $U(r)$
($U^{(aux)}(r)$). Solutions of the eqs. (\ref{ethird})
describe the decay of quasi-bound state ${\Phi}_i$~ in the continuum.
Reduced matrix elements of the source term involve products of two
annihilation operators and one creation operator :
${\cal R}^{j_{\alpha}}_{{\gamma}{\delta}(L){\beta}}=
(a^\dagger_{\beta}({\tilde a}_{\gamma}{\tilde a}_{\delta})^L)^{j_{\alpha}}$.
In the SMEC calculations for $^{17}$F and $^{17}$O, 
the matrix elements of the source term depend
sensitively on the percentage of the shell closure in $^{16}$O, {\it i.e.}, on
the amount of correlations both in the g.s. of
 $^{16}$O and in the considered states of $^{17}\mbox{F}$ or $^{17}$O 
\cite{bnop4}. The total wave function is expressed by three functions:
${\Phi}_i$~, ${\xi}_{E}^{c}$~ and ${\omega}_i$ \cite{bartz3,bnop1,bnop2} :
\begin{eqnarray}
\label{eq2}
{\Psi}_{E}^{c} = {\xi}_{E}^{c} + \sum_{i,j}({\Phi}_i + {\omega}_i)
\frac{1}{E - H_{QQ}^{eff}}
\langle {\Phi}_{j}\mid H_{QP} \mid{\xi}_{E}^{c}\rangle ~ 
\end{eqnarray}
where : 
\begin{eqnarray}
H_{QQ}^{eff}(E) = H_{QQ} + H_{QP}G_{P}^{(+)}H_{PQ} ~ \ ,
\end{eqnarray}
is the energy dependent effective
SM Hamiltonian in $Q$ subspace which contains couplings to the continuum.
Operator $H_{QQ}^{eff}(E)$ is hermitian for
energies below the particle emission threshold and non-hermitian 
for energies higher than the threshold. 
The eigenvalues ${\tilde {E_i}} - \frac{1}{2}i{\tilde {{\Gamma}_i} }$ are
complex for decaying states and
depend on the energy $E$ of the particle in the continuum.
The energy and the width of resonance states are determined by the condition:
$\tilde{E_i}(E) = E$. The eigenstates
corresponding to these eigenvalues can be obtained by the orthogonal but in
general non-unitary transformation \cite{bnop1,bnop2,weid,bartz3}.
Inserting them in (\ref{eq2}), one obtains symmetrically 
the new continuum many-body 
wave function modified by the discrete states : 
\begin{eqnarray}
\label{cons}
{\Psi}_{E}^{c} = {\xi}_{E}^{c} + \sum_{i}^{}{\tilde {\Omega}_i}
\frac{1}{E - {\tilde E_i}
+ (i/2){\tilde {\Gamma}_i}} \langle 
{\tilde {\Phi}_i} \mid H \mid {\xi}_{E}^{c}\rangle ~ \ ,
\end{eqnarray}
and the new
discrete state wave function modified by the coupling 
to the continuum states:
\begin{eqnarray}
\label{diss}
{\tilde {\Omega}_i} = {\tilde {\Phi}_i} + \sum_{c}
\int_{{\varepsilon}_c}^{\infty} dE^{'} {\xi}_{E^{'}}^{c}
\frac{1}{E^{(+)} - E^{'}}
\langle {\xi}_{E^{'}}^{c}\mid H \mid {\tilde {\Phi}_i}\rangle ~ \ .
\end{eqnarray}
These SMEC wave functions will be used in this paper to calculate spectra of
$^{17}$F, $^{17}$O, the first-forbidden $\beta -$decays : $^{17}$Ne($\beta^+)^{17}$F, 
$^{17}$N($\beta^-)^{17}$O, and the radiative proton capture 
reaction $^{16}$O$(p,\gamma )^{17}$F.

Figs. 1 and 2 show SMEC energies and widths  for
positive parity (l.h.s.\ of the plot) and negative parity (r.h.s.\ of the
plot) states of $^{17}\mbox{F}$ and $^{17}$O, respectively.    
Large breaking of mirror symmetry, which can be seen by comparing spectra in
Figs. 1 and 2, is due to the combined 
effect of the low separation energy and the Coulomb field.
The continuum coupling, which due to different positions of the lowest particle emission 
thresholds acts differently in $^{17}$F and $^{17}$O, cannot fully account for the observed mirror 
symmetry breaking. The simplest way of correcting this deficiency is to
adjust the spacing of $d_{5/2}$ and $s_{1/2}$ s.p. orbitals
in ZBM interaction in such a way that the experimental energy splitting 
between the g.s. $5/2_1^{+}$ and the first excited state 
$1/2_1^{+}$  in $^{17}$O and $^{17}$F 
is reproduced by the SMEC calculation. In this way, 
the s.p. energy of $d_{5/2}$ in
ZBM interaction becomes : $\varepsilon_{d_{5/2}}=3.95$ and 3.5  
in $^{17}$F and $^{17}$O, respectively. The energy of $s_{1/2}$ orbital remains 
$\varepsilon_{s_{1/2}}=3.3$ in both cases. In the following,
these two hybrid interactions will be called ZBM-F and ZBM-O interactions
, respectively. The Thomas-Ehrmann shift is then
taken into account through the combined effect of the mirror symmetry breaking
continuum coupling and the modification of the s.p. energies of the effective
SM interaction.

The coupling matrix elements between the $J^{\pi}=0_{1}^{+}$ 
g.s.\ wave function of $^{16}\mbox{O}$ and all considered states 
in $^{17}\mbox{F}$ and $^{17}$O are calculated using the
density dependent DDSM1 interaction \cite{bnop3,bnop4}. 
The coupling to the continuum states is given 
by the matrix elements of ${\cal R}^{j_{\alpha}}_{{\gamma}{\delta}(L){\beta}}$ 
between g.s. of $^{16}$O and all considered 
states in $^{17}$F and $^{17}$O. 
The calculation of the radial wave functions and radial formfactors of the
coupling to the continuum states goes as follows. 
The s.p.\ wave functions, which in the many body states $J^{\pi}$ of
$^{17}$F are not modified by the selfconsistent
correction to the finite initial potential, are calculated using the 
auxiliary  potential $U^{(aux)}$. This potential, which contains central
, spin-orbit and Coulomb parts, 
is adjusted to yield the binding energies of proton
s.p. orbits $0d_{5/2}$ and $1s_{1/2}$ at the 
experimental binding energies of $5/2_{1}^{+}$ and $1/2_{1}^{+}$ states 
in $^{17}$F. This adjustement of binding energies of s.p. orbits makes sense because the
many-body wave function $5/2_1^+$ (respectively $1/2_1^+$) has a large amplitude
of the component with one particle in $0d_{5/2}$ (respectively $1s_{1/2}$) s.p.
state outside of the $^{16}$O core. The parameters of this potential are given in Table 1 for
different values of the diffuseness parameter $a$. Without Coulomb term, 
the same potential is used also to calculate radial formfactors for 
neutrons in $^{17}$F. In the calculation for
$^{17}$O, again the same potential with the Coulomb term of $^{17}$O 
is used. Such a potential yields binding energies 
of neutron $0d_{5/2}$ and $1s_{1/2}$ s.p. orbits very closely to the 
experimental binding energies of $5/2_{1}^{+}$ and $1/2_{1}^{+}$ states 
in $^{17}$O. If the coupling between bound and scattering states 
modifies the s.p. wave function $\phi_{l,j}$ in the many body state $J^{\pi}$, 
then the depth of the initial potential $U(r)$ is readjusted to ensure
that the energy of the s.p. state $\phi_{l,j}$ 
in the selfconsistent potential $U^{(sc)}(r)$ is the same as 
the energy of this state in the auxiliary potential $U^{(aux)}$. 
 The remaining parameters : $R_0$, $a$, $V_{SO}$, of the initial potential 
are the same as in $U^{(aux)}$. This procedure for radial wave functions
yields the same asymptotic property for a given s.p wave function in all
different channels.  

The calculation of the first-forbidden $\beta -$decays presented in this work
are the extension of the calculations of
Towner and Hardy \cite{towner} (see also Ref. \cite{millener1} for the
presentation of the method). In the following we give only
few elements of this approach to introduce the notation used in this work. 
The calculation of the absolute decay rate uses 
\cite{scho66,wil77} : $ft_{1/2}=6170$ s, where $t_{1/2}$ is the partial
half-life and :
\begin{eqnarray}
\label{eq1}
f=\int_1^{W_0}C(W)F(Z,W)(W^2-1)^{1/2}W(W_0-W)^2dW  ~ \ .
\end{eqnarray}
In the above expression, $W$ is the $\beta -$energy, $W_0$ is the maximum
$\beta -$energy and $Z$ is the charge of the final nucleus. The
first-forbidden shape factor $C(W)$ can be written to a good approximation as
\cite{scho66,towner} : $C(W)=k(1+aW+bW^{-1}+cW^2)$, where coefficients $k$,
$a$, $b$ and $c$ depend on the nuclear matrix elements, $W_0$ and $\xi
=Ze^2/2R$. (For the nuclear radius $R$ we use the prescription of Wilkinson
\cite{wil77}.) Consequently :
\begin{eqnarray}
\label{eq2a}
f=k(I_0+aI_1+bI_{-1}+cI_2) ~ \ .
\end{eqnarray}                  
The evaluation of integrals $I_k$ is given in
\cite{towner1}. Relation of the coefficients $k, a, b, c$ 
to the nuclear matrix elements of different rank is \cite{scho66,behr71} :
\begin{eqnarray}
\label{eq9}
k&=&\left[\zeta_0^2+\frac{1}{9}w^2\right]^{(0)}+\nonumber \\
&+&\left[\zeta_1^2+\frac{1}{9}(x+u)^2-\frac{4}{9}\mu_1\gamma_1u(x+u)+\frac{1}{18}W_0^2(2x+u)^2-
\frac{1}{18}\lambda_2(2x-u)^2\right]^{(1)}+\nonumber \\
&+&\frac{1}{12}\left[z^2(W_0^2-\lambda_2)\right]^{(2)} ~ \ , 
\nonumber                                 \\
ka&=&-\left[\frac{4}{3}uY-\frac{1}{9}W_0(4x^2+5u^2)\right]^{(1)}-\left[\frac{1}{6}z^2W_0\right]^{(2)}
~ \ , \\
kb&=&\frac{2}{3}\mu_1\gamma_1\{ -[\zeta_0w]^{(0)}+[\zeta_1(x+u)]^{(1)}\} ~ \ , \nonumber  \\
\nonumber
kc&=&\frac{1}{18}[8u^2+(2x+u)^2+\lambda_2(2x-u)^2]^{(1)}+\left[\frac{1}{12}z^2(1+\lambda_2)\right]^{(2)}
~ \ , \nonumber 
\end{eqnarray}
where
\begin{eqnarray}
V&=&\xi^{'}v+\xi w^{'} ~ \ , \hspace{3.1cm} \zeta_0=V+\frac{1}{3}wW_0 ~ \ , \nonumber
\\
Y&=&\xi^{'}y-\xi (u^{'}+x^{'}) ~ \ , \hspace{2cm} 
\zeta_1=Y+\frac{1}{3}(u-x)W_0 ~ \ .
\nonumber
\end{eqnarray}
The quantities $\mu_1, \gamma_1$ and $\lambda_2$ are defined in terms of the
electron wave functions and have values close to unity \cite{scho66}. The
non-relativistic form of the nuclear matrix elements is \cite{behr71} :
\begin{eqnarray}
\label{eq3}
w&=&\lambda\sqrt{3}\langle J_fT_f|||ir[{\bf C}_1\times{\bf {\sigma}}]^0{\bf
{\tau}}|||J_iT_i\rangle C ~ \ , \nonumber \\     
x&=&-\langle J_fT_f|||ir{\bf C}_1\cdot{\bf {\tau}}|||J_iT_i\rangle C ~ \ , \\     
u&=&\lambda\sqrt{2}\langle J_fT_f|||ir[{\bf C}_1\times{\bf {\sigma}}]^1{\bf
{\tau}}|||J_iT_i\rangle C ~ \ , \nonumber        \\
z&=&-2\lambda\langle J_fT_f|||ir[{\bf C}_1\times{\bf {\sigma}}]^2
{\bf {\tau}}|||J_iT_i\rangle C ~ \ , \nonumber   
\end{eqnarray}
where $\lambda =-g_A/g_V=1.2599(25)$ \cite{towner2} and :
\begin{eqnarray}
C=\frac{\langle T_iT_{z_i}1\pm 1|T_fT_{z_f}\rangle}{[2(2J_i+1)(2T_f+1)]^{1/2}}
\end{eqnarray}
The remaining matrix elements in the non-relativistic form are :
\begin{eqnarray}
\label{eq4}
\xi^{'}v&=&-\lambda\sqrt{3}\langle J_fT_f|||\frac{i}{M}[{\bf {\sigma}}\times{\bf
{\nabla}}]^0{\bf {\tau}}|||J_i T_i\rangle C ~ \ , \nonumber \\
\xi^{'}y&=&\langle J_fT_f|||\frac{i}{M}{\bf {\nabla}}\cdot{\bf {\tau}}|||J_iT_i\rangle C ~
\ , 
\end{eqnarray}
where $M$ is the nucleon mass. It has been found by Warburton et al.
\cite{war91,war94} that the matrix element $\xi^{'}v$ of the time-like piece of
the axial current is strongly enhanced by meson-exchange currents, mainly the
one-pion exchange. The enhancement factor that multiplies the
impulse-approximation axial-charge matrix element, has been determined by
comparison to experiment for $A\sim 16$ to be $\epsilon=1.61\pm 0.03$. In all
calculations discussed in this work, we multiply the matrix element $\xi^{'}v$
by a constant factor 1.61, like
in the analysis of Borge et al. \cite{borge}. Somewhat smaller enhancement
factor has been used in the study of Millener \cite{millener}. This meson
exchange contribution to the axial charge is very important, changing the
absolute decay rates of the first-forbidden decays $f^{-}$ and $f^{+}$ by a
factor $\sim 3.7$ (see also the discussion in \cite{millener}). However, 
its influence on the ratio $f^{+}/f^{-}$ is somewhat less important.

The matrix elements $w^{'}, x^{'}$ and $u^{'}$ 
are obtained from $w,x$ and $u$ by
including an extra factor in the radial integral \cite{behr71}.
For the higher energy part of the spectrum in light nuclei, the term $kbW^{-1}$
can be neglected and the energy dependence of $C(W)$ is determined entirely by
the matrix elements of multipolarity 1 and 2 \cite{towner}, {\it i.e.} by $ka$
and $kc$. As written above,
the expressions for the matrix elements apply to electron emission. For
positron emission we have to make the following replacement : $Z \rightarrow
-Z$ and $\lambda \rightarrow -\lambda$. Even assuming perfect mirror symmetry,
the nuclear matrix elements combine in $f$ with different signs what gives a
different shape correction factor for electron and positron decays. 
The evaluation of matrix elements in
the basis of selfconsistently determined s.p. wave functions in the initial
$J_i^{\pi}=1/2^-$ and final $J_f^{\pi}=1/2^+$ many body states 
follows the procedure described by
Towner and Hardy \cite{towner} and adopted in Ref. \cite{borge}. 

\section{Discussion}
An exceptionally large asymmetry of mirror $\beta -$decays : 
$^{17}$N($J^{\pi}=1/2_1^-$)$\longrightarrow ^{17}$O($J^{\pi}=1/2_1^+$) and
$^{17}$Ne($J^{\pi}=1/2_1^-$) $\longrightarrow ^{17}$F($J^{\pi}=1/2_1^+$),  
raises the question about a role of largely different radial sizes
of $1s_{1/2}$ s.p. wave functions in the initial state $1/2_1^-$ of $^{17}$Ne/$^{17}$N. In the final state $1/2_{1}^+$ 
, the large asymmetry of radial configurations of $^{17}$F, $^{17}$O and the
halo structure of the $^{17}$F is well known. A dominant component of the
$1/2_1^-$ wavefunction in $^{17}$Ne/$^{17}$N is
$\pi(0d_{5/2}^2)\nu(0p_{1/2}^{-1})$/$\nu(0d_{5/2}^2)\pi(0p_{1/2}^{-1})$
configuration outside the core of $^{16}$O but this configuration does not play
a role in the first-forbidden $\beta-$decay. The dominant contribution comes
from the small component of the wave function
$\pi(1s_{1/2}^2)\nu(0p_{1/2}^{-1})$/$\pi(1s_{1/2}^2)\nu(0p_{1/2}^{-1})$ in the
g.s. of $^{17}$Ne/$^{17}$N. For this
component, $1s_{1/2}$ nucleon makes a
transition to fill the $0p_{1/2}$ hole.

Assuming strict mirror symmetry in the $\beta -$decays, the SM analysis
yields the ratio $f^+/f^-$ which is much smaller ($f^+/f^-\simeq 9.6$) 
than the experimental value ($24\pm 4$) \cite{borge}. 
Allowing for the variation of radii of
selected s.p. orbits, Borge et al. \cite{borge} have shown 
that the observed values of the ratio $f^{+}/f^{-}$ 
can be reproduced in the SM analysis assuming a difference 
of $\sim 0.6$ fm between the oscillator length parameters 
for proton and neutron $1s_{1/2}$ orbitals. 
It was assumed that radii of $1s_{1/2}$ neutron orbits in
$^{17}$N and $^{17}$O are the same. Similar assumption has been made for radii of $1s_{1/2}$ proton
orbits in $^{17}$Ne and in $^{17}$F. 
Remaining s.p. orbits have been assumed to have the same radius in 
$^{17}$F, $^{17}$Ne, $^{17}$O and $^{17}$N. 

High sensitivity of the $\beta -$decay asymmetry 
to the spatial features of s.p. orbits \cite{borge},
provides a challenge for the SMEC approach because 
radial characteristics for certain 
orbits as well as their asymptotic properties 
are determined consistently for each studied nucleus 
and, moreover, model parameters 
( the effective SM interaction, the residual coupling, radius/diffuseness 
of the initial/auxiliary average potential) are determined by analyzing 
different reaction and spectroscopic data in the same many-body framework.  

In Fig. 3 we show $f^+$, $f^-$ and $f^+/f^-$, which are calculated  in SMEC 
for different values of the diffuseness parameter of the initial potential. In
each case, the same diffuseness is taken for initial potentials in $^{17}$Ne,
$^{17}$N, $^{17}$F and $^{17}$O. The shaded areas
show the experimental limits for these values. $f^+$ and $f^-$ are calculated
with ZBM-F and ZBM-O interactions (the solid lines), respectively. The dotted
line in the middle part of Fig. 3 
shows results for $f^-$ obtained with ZBM-F interaction. The
corresponding ratio $f^+/f^-$, shown by the dotted line in the lower part of
Fig. 3, corresponds then to strictly mirror symmetric SM interaction. The coupling to continuum is
given by DDSM1 residual interaction in all studied cases.
The calculations of the radial wave functions and radial formfactors of the
coupling to the continuum states for $^{17}$F/$^{17}$O have been described
above. In the calculations for
$^{17}$Ne/$^{17}$N, we employed the initial/auxiliary potentials obtained from
those for $^{17}$F/$^{17}$O by an appropriate modification of the
Coulomb potential. 
One should mention, that initial potentials $U(r)$ ($U^{(aux)}(r)$) 
for different diffuseness parameters, yield very similar spectra for each considered
nucleus. Therefore, 
the energy spectra do not provide any constraint on 
the diffuseness of the average field. 
The experimental value of the ratio $f^+/f^-$ can be reproduced using 
$U(r)$ ($U^{(aux)}(r)$) with a very large value of the diffuseness parameter 
($a \sim 0.8$ fm). However, none of these potentials with extremely thick
skin can reproduce the experimental values for either $f^+$ ($f^+_{exp}=927\pm 95$) 
or $f^-$ ($f^-_{exp}=44\pm 7$). 

The overall contribution of matrix elements of rank 1 in $f^+$ and $f^-$ increases with
increasing diffuseness parameter from $\sim 11\%$ at $a=0.4$ fm to $\sim 21\%$
at $a = 0.8$ fm. This strong relative increase is similar for $f^+$ and
$f^-$. More insight into the $a-$dependence of $f^+$ and $f^-$ can be gained from 
Fig. 4 (see also Tables 2 and 3) 
which shows the matrix elements : 
$kI_0$ (the upper part) and the sum $kaI_1+kcI_2$, separately for
$^{17}$Ne($\beta^+$)$^{17}$F (the solid lines) and 
$^{17}$N($\beta^-$)$^{17}$O (the dotted lines) decays. $kI_0$ is the
combination of nuclear matrix elements of rank 0 and 1. Coefficient $k$ is the
energy-independent term of the first-forbidden shape factor $C(W)$. The sum of
$kaI_1$ and $kcI_2$ depends on nuclear matrix elements of rank 1 only. 
In our case, the nuclear matrix elements of rank 2
associated with the operator $z^2$ ({\it c.f.} equation (\ref{eq9})) are identically equal zero. 
The coefficients $ka$ and $kc$ determine $W-$ and $W^2-$dependence of the shape
correction factor $C(W)$.
ZBM-F and ZBM-O interactions are used in the calculation of $\beta^+$ and $\beta^-$ decays
respectively. $(kI_0)^+$ and $(kI_0)^-$ decrease 
with increasing $a$ and the decrease rate is similar in both cases. 
The strong increase of the ratio $f^+/f^-$ for large $a$ is caused by 
a different $a-$dependence of $(kaI_1+kcI_2)^+$ and $(kaI_1+kcI_2)^-$. 
With increasing diffuseness of $U(r)$ ($U^{(aux)}(r)$), 
$(kaI_1+kcI_2)^+$ decreases strongly whereas $(kaI_1+kcI_2)^-$ 
remains approximately constant and is negative. In the
studied range of $a$ values, the ratio $(kI_0)^+/(kI_0)^-$ increases only 
by $\sim 25\%$ whereas 
$[(kI_0)^++(kaI_1+kcI_2)^+]/[(kI_0)^-+(kaI_1+kcI_2)^-]\simeq f^+/f^-$ 
increases by $\sim 70\%$. At the same time,
$kbI_{-1}$ is small and almost constant. 
Hence, those nuclear matrix elements of multipolarity 1 which determine the
$\beta-$energy dependence of the first-forbidden shape factor play also a salient
role in exhibiting the asymmetry in the mirror first-forbidden $\beta-$decays :
$^{17}$Ne($\beta^+$)$^{17}$F and $^{17}$N($\beta^-$)$^{17}$O. 

This effect is independent
of whether ZBM-F or ZBM-O effective interaction is used 
in describing $\beta^-$ decay : $^{17}$N($\beta^-$)$^{17}$O. 
The dashed line in Fig. 4 shows the dependence of
$(kI_0)^-$ and $(kaI_1+kcI_2)^-$ on the parameter $a$ of the first guess
potential $U(r)$, which is calculated using the  
ZBM-F interaction, as in the calculation of $^{17}$Ne($\beta^+$)$^{17}$F decay
rate. As before, $(kI_0)^-$ decreases strongly and $(kaI_1+kcI_2)^-$ 
remains constant. This shows that the $a-$dependence of the ratio
$f^+/f^-$ {\em is not related} either to the choice of the particular variant of ZBM
interaction or to the assumption of strict mirror symmetry in $\beta^+-$ and
$\beta^--$ decays. On the other hand, the overall magnitude of
this ratio depends strongly on the amount of mirror symmetry breaking in the
many-body wave function. 
Replacing ZBM-O by ZBM-F in the calculation of $^{17}$N($\beta^-$)$^{17}$O
decay rate, leads to the decrease of $f^-$ by a factor $\sim 1.8$ 
(see also Fig. 4). This sensitivity, which has been stressed by Millener
\cite{millener}, is stronger than the $a-$dependence of $f^-$.
The separate contributions of $(kaI_1)^{\pm}$, $(kbI_{-1})^{\pm}$, 
$(kcI_2)^{\pm}$ are given in Table 2. Nuclear matrix elements are shown in
Table 3 for different SM interactions and SMEC radial wave functions. 
$f^-$ is calculated for
both ZBM-O and ZBM-F interactions. One can see that  
in the considered range of surface diffuseness parameters the enhancement rate
of the ratio $f^+/f^-$ is $\sim 1.4$, independently of the assumption of strict 
mirror symmetry of the SM interaction. 

In order to quantify the relative importance of radial sizes of the 
s.p. states and the mirror symmetry breaking in the many body wave functions, 
one has to constrain the surface diffuseness of the initial potential.
Such a constraint can be provided by the proton radiative capture
cross section $^{16}$O$(p,\gamma )^{17}$F to the 'proton halo state'
$1/2_1^+$, which is very sensitive to the size of $1s_{1/2}$ 
proton s.p. wave function \cite{bnop4,rolfs}. The astrophysical $S$-factor
as a function of the c.m.\ energy is shown in Fig. 5 
for three values of the diffuseness parameter of the initial potential : 
$a=0.4$ fm (the dotted line), 
0.55 fm (the solid line) and 0.8 fm (the dashed line), separately for each 
decay branch : 
$^{16}\mbox{O}(p,\gamma)^{17}\mbox{F}(J^{\pi}=1/2_1^{+})$ and 
$^{16}\mbox{O}(p,\gamma)^{17}\mbox{F}(J^{\pi}=5/2_1^{+})$. 
One should notice that in all cases, energies of proton s.p. orbits $0d_{5/2}$
and $1s_{1/2}$ in the selfconsistent potentials for diffenent $a$ are at the experimental
binding energies of $5/2_1^+$ and $1/2_1^+$ states in $^{17}$F, respectively. The sum of contributions
from these two decay branches is shown in the lower part of Fig. 5.
The calculations are performed using SMEC
wave functions (5) and (6) which have been obtained exactly in the same way and
for exactly the same input parameters as those used in the calculation of both the
decay rate $^{17}$Ne($\beta^+$)$^{17}$F (see Figs. 3 and 4) 
and the spectrum of $^{17}$F (Fig. 1). 
The scale of excitation energy is the same as c.m. energy in the $p +
^{16}\mbox{O}$ system. The photon energy is given by the difference of
c.m.\ energy of $[^{16}\mbox{O} + \mbox{p}]_{J_{i}^{\pi}}$ system
and the experimental energy of the final state ($1/2_1^+$ or $5/2_1^+$) 
in $^{17}\mbox{F}$. 
We have taken into account all possible $E1$, $E2$, and $M1$ 
transitions from incoming $s$, $p$, $d$, $f$, and $g$ waves. It is clearly seen in Fig. 5 
that the SMEC calculation for
$a=0.4$ fm underestimates the experimental capture cross-section. 
On the other hand, the
calculation for $a=0.8$ fm, for which the ratio $f^+/f^-$ agrees with the data (see
Fig. 3), overestimates  strongly the data. 
Realistic values of the surface diffuseness parameter, which are compatible with
the proton capture data \cite{morlock}, are : $a=0.55\pm 0.05$ fm. 
In this range, $f^+$ agrees perfectly with the experimental data, whereas
$f^-$ is too big. For $a=0.55$ fm, $f^-$ 
overshoots the data by a factor $\simeq 1.4$.

Since the radial dependences which
are consistent with the proton capture reaction data, give excellent fit of
both the $\beta^+-$decay rate and the spectrum of $^{17}$F, the discrepancy found for 
$f^-$ and $f^+/f^-$ should be explained by the deficiency of the effective SM
interaction to reproduce the configuration mixing in $^{17}$O  ($^{17}$N).
If one includes the effect of mirror symmetry breaking 
through the modification of energies of s.p. orbitals of the SM, 
then the experimental value of $f^-$ can be reproduced by SMEC
wave function for $a=0.55$ fm using 
: $\varepsilon_{d_{5/2}}=3.21$ ($\varepsilon_{s_{1/2}}=3.3$). 
For this interaction, called ZBM-O$^*$, Fig. 6 shows SMEC energies and
widths  for positive parity (l.h.s.\ of the plot) and negative parity (r.h.s.\ of the
plot) states of $^{17}$O. Negative parity states are reproduced better 
by ZBM-O* than by ZBM-O SMEC calculations ({\it c.f.} Fig. 2 and Fig. 6). The splitting   
of $5/2_1^+$ and $1/2_1^+$ states in SMEC/ZBM-O* is slightly larger than the experimental
splitting. The spacing $\varepsilon_{d_{5/2}} - \varepsilon_{s_{1/2}}$ in
ZBM-O* is reduced too much as compared to ZBM-F interaction
to account for the absence of the expected renormalization of the
two-body matrix elements. Since the dominant configuration
in $5/2_1^+$ and $1/2_1^+$ states is the 1p-0h component outside a
closed core of $^{16}$O with one nucleon in either $d_{5/2}$ or $s_{1/2}$
shells, therefore it is quite natural that the
splitting $\varepsilon_{d_{5/2}} - \varepsilon_{s_{1/2}}$ in ZBM
interaction has a particularly strong effect on the relative position of these
two states. 

An interesting supplementary quantity is the $B(E2)$ transition matrix
element between $1/2_1^+$ and $5/2_1^+$ bound states of $^{17}$F and $^{17}$O. 
Assuming the effective charges :
$e_p \equiv 1+\delta_p$, $e_n \equiv \delta_n$, with the polarization charge 
$\delta =\delta_p=\delta_n=0.2$, which are suggested by 
the theoretical estimates \cite{kirson}, one finds in SMEC/ZBM-F 
with $a=0.55$ fm : $B(E2)=74.85$ e$^2$fm$^4$ for $^{17}$F. 
The experimental value for this transition is $B(E2)_{exp}=64.92$ e$^2$fm$^4$.
In $^{17}$O one finds similar results : $B(E2)=3$ and 3.2 e$^2$fm$^4$ in SMEC/ZBM-O and
SMEC/ZBM-O* calculations, respectively. 
The experimental value for this transition is $B(E2)_{exp}=6.2$ e$^2$fm$^4$.

Fig. 7 shows $f^+$, $f^-$ and $f^+/f^-$, which are calculated  
in SMEC for different values of the diffuseness parameter of the 
initial potential. The shaded areas
give the experimental limits. $f^+$ and $f^-$ are calculated
with ZBM-F and ZBM-O* interactions, respectively. 
An essential physical parameter is the
amplitude of a component $(1s_{1/2}^20p_{1/2}^{-1})$ in the g.s.
wave functions of $^{17}$Ne and $^{17}$N. These amplitudes are shown in 
Table 4 together with the dominant 1p-0h component of the wave function in the
final state $1/2_1^+$. One can notice that the amplitude of
$(1s_{1/2}^20p_{1/2}^{-1})$ configuration in $^{17}$Ne (ZBM-F) is $\sim 30 \%$ bigger
than in $^{17}$N (ZBM-O*). One should however recall that this important
difference concerns the small component of the wave function and the large
component of the many-body wave function in the g.s. of $^{17}$Ne and $^{17}$N
differs by less than $5 \%$. The same change of the effective interaction induces
only a small ($\sim 12 \%$) difference of the dominant 1p-0h configuration 
in $1/2_1^+$ wave function of $^{17}$F and $^{17}$O.
                                 
In conclusion, we have found that the increase of the ratio $f^+/f^-$ is
correlated with the increase of radius of weakly bound $1s_{1/2}$ s.p. orbit in $^{17}$F, 
in accordance with the conclusion of Borge et al. \cite{borge}, but to obtain 
agreement with the data one has to assume an
unrealistic geometry of the self-consistent potentials in $^{17}$Ne, $^{17}$N,
$^{17}$F and $^{17}$O, which disagrees with the proton capture data. 
If, consistently, one takes into account the constraint from the
capture data, one can estimate mirror symmetry breaking in the
SM effective interaction in ZBM space for $A=17$ nuclei. This effect is less
than $\sim 5 \%$ for the dominant term in the g.s. wave function and about
$\sim 30 \%$ for the small component of the wave function which plays a crucial
role in the first-forbidden $\beta-$decay. In the final nucleus, the mirror
symmetry breaking is less than $\sim 12 \%$. These
estimates are expected to depend somewhat on the SM space used. In particular,
the absence of $0p_{3/2}$ and
$0d_{3/2}$ subshells in the ZBM space may lead to the amplification of
the sensitivity in the $1s_{1/2} \rightarrow 0p_{1/2}$ contribution to the
charge-dependent effects \cite{millener}.

Many nuclear
matrix elements contribute to the transition probability so the experimental
determination of the lifetime and spectrum shape alone is usually insufficient
to determine them all. A unique first-forbidden $\beta-$transition from
$^{17}$N to the g.s. of $^{17}$O is known ($f=24\pm 8$) \cite{silbert}. Unfortunately, this transition tests
the nuclear matrix element of rank 2 which are absent in the first-forbidden
$\beta-$transition and which cannot be calculated reliably in a small effective
SM space. SMEC gives for this unique first-forbidden transition 
a value which is $\sim 3$ times larger than the experimental
value and depends weakly on the chosen hybrid of the ZBM force. This
discrepancy is expected to be solved in calculations using large basis up to
4$\hbar\omega$ \cite{war94,war92}.
Only in favourable circumstances, the nuclear structure information can be unambiguously extracted from 
the first-forbidden $\beta -$decays and relative importance of the
configuration mixing (internal mixing) and the exotic radial dependences 
of s.p. wave functions (external mixing) can
be disentangled. It seems that the first-forbidden $\beta -$decays, which
depend sensitively both on the fine details of the 
SM effective interaction and on
the radial formfactors of the wave functions in mirror systems will be
particularly difficult to exploit as a direct and unambiguous 
source of information about unstable nuclei.

\vskip 1truecm

{\bf Acknowledgments}\\
We thank the Institute for Nuclear Theory at the University of Washington for
its hospitality and the Department of Energy for partial support during the
completion of this work. It is a pleasure to thank G. Mart\'{\i}nez-Pinedo for
numerous discussions and to K. Langanke for valuable suggestions. 
This work was partly supported by
KBN Grant No. 2 P03B 097 16 and the Grant No. 76044
of the French - Polish Cooperation.

\vfill
\newpage

\begin{table}[h]

\caption{The parameters of initial potentials $U(r)$ 
used in the calculations of self-consistent potentials $U^{(sc)}(r)$ for
$s_{1/2}$ and $d_{5/2}$ s.p. wave functions
in $1/2_1^+$ weakly bound state and $5/2^+$ ground state of $^{17}$F,
respectively. The residual
coupling of $Q$ and $P$ subspaces is given by the DDSM1 interaction.
For all considered cases the radius of the potential is $R_0=3.214\,$fm. 
For more details, see the description in the text.}
\label{tab1}
\begin{center}
\begin{tabular}{|c|c|c|}
\hline
Diffuseness [fm]& $V_0$ [MeV] & $V_{SO}$ [MeV]\\
\hline
0.40 & $ -55.119 $ & $ -1.383 $ \\
0.45 &  $ -54.432 $ & $ 0.097 $ \\
0.50 &  $ -53.6975 $ & $ 1.5185 $ \\
0.55 &  $ -52.929 $ & $ 2.886 $ \\
0.60 &  $ -52.139 $ & $ 4.203 $ \\
0.65 &  $ -51.334 $ & $ 5.475 $ \\
0.70 &  $ -50.521 $ & $ 6.706 $ \\
0.75 &  $ -49.705 $ & $ 7.898 $ \\
0.80 &  $ -48.889 $ & $ 9.054 $ \\
0.85 &  $ -48.076 $ & $ 10.178 $ \\
\hline
 \end{tabular}
\end{center}
\end{table}

\vfill
\newpage

\begin{table}[h]

\caption{Nuclear matrix elements and decay rates are shown for different values of the
diffuseness parameter of the 'first guess', initial 
potential $U(r)$. Quantities for $\beta^+-$decays are
calculated using ZBM-F interaction. For $\beta^--$decay, ZBM-O, ZBM-O* and
ZBM-F interactions are used. 
}
\label{tab2}
\begin{center}
\begin{tabular}{|c|c|c|c|c|c|c|}
\hline
Interaction & Matrix element & $a=0.4$ fm & $a=0.5$ fm & $a=0.6$ fm & $a=0.7$ fm & $a=0.8$ fm \\
\hline
ZBM-F& $(kI_0)^+$ &1024.88 & 872.67 & 718.37 & 584.54 & 483.76 \\
 & $kaI_1^+$  & -5.10 & -31.53 & -72.23 & -129.54 & -205.88 \\
 & $kbI_{-1}^+$  & 1.60 & 1.62 & 1.67 & 1.76 & 1.88 \\
 & $kcI_2^+$  & 112.26 & 124.77 & 145.90 & 176.54 & 217.50 \\
 & ($kaI_1^+$+$kcI_2^+$) & 107.15 & 93.24 & 73.66 & 47.00 & 11.62 \\
 & $f^+$  & 1133.64 & 967.53 & 793.70 & 633.30 & 497.26 \\
\hline
ZBM-O & $(kI_0)^-$ & 94.23 & 80.48 & 65.27 & 50.99 & 39.11 \\
 & $kaI_1^-$  & -17.68 & -18.32 & -19.46 & -21.13 & -23.32 \\
 & $kbI_{-1}^-$ & 0.39 & 0.38 & 0.38 & 0.37 & 0.36 \\
 & $kcI_2^-$  & 4.46 & 5.04 & 5.99 & 7.35 & 9.185 \\
 & ($kaI_1^-$+$kcI_2^-$) & -13.28 & -13.1 & -13.47 & -13.76 & -14.13 \\
 & $f^-$   & 81.39 & 67.58 & 52.17 & 37.59 & 25.34 \\
\hline
ZBM-O* & $(kI_0)^-$ & 66.01 & 56.355 & 45.69 & 35.72 & 27.5 \\
 & $kaI_1^-$  & -13.475 & -13.96 & -14.83 & -16.09 & -17.755 \\
 & $kbI_{-1}^-$ & 0.29 & 0.285 & 0.28 & 0.27 & 0.26 \\
 & $kcI_2^-$  & 3.39 & 3.83 & 4.55 & 5.59 & 6.97 \\
 & ($kaI_1^-$+$kcI_2^-$) & -10.085 & -10.13 & -10.28 & -10.5 & -10.78 \\
 & $f^-$   & 56.215 & 46.51 & 35.69 & 25.49 & 16.98 \\
\hline
ZBM-F & $(kI_0)^-$ & 165.015 & 141.02 & 114.46 & 89.40 & 68.36 \\
 & $kaI_1^-$   & -27.94 & -28.96 & -30.77 & -33.40 & -36.88 \\
 & $kbI_{-1}^-$  & 0.64 & 0.63 & 0.62 & 0.605 & 0.585 \\
 & $kcI_2^-$   & 7.08 & 8.00 & 9.51 & 11.68 & 14.58 \\
 & ($kaI_1^-$+$kcI_2^-$)  & -20.86 & -20.95 & -21.25 & -21.72 & -22.29 \\
 & $f^-$  & 144.80 & 120.70 & 93.82 & 68.29 & 46.65 \\
\hline
 \end{tabular}
\end{center}
\end{table}

\vfill
\newpage

\begin{table}[h]

\caption{Nuclear matrix elements for different values of the
diffuseness parameter of the initial potential $U(r)$. 
Quantities for $\beta^+-$decays are
calculated using ZBM-F interaction. For $\beta^--$decay, ZBM-O, ZBM-O* and
ZBM-F interactions are used. 
}
\label{tab3}
\begin{center}
\begin{tabular}{|c|c|c|c|c|c|c|}
\hline
Quantity  & Matrix element & $a=0.4$& $a=0.5$& $a=0.6$&$a=0.7$ & $a=0.8$ \\
\hline
$f^+$(ZBM-F) & ${\xi}^{'}v$ &24.05 & 22.62 & 21.03 & 19.43 & 17.89      \\
             & ${\xi}^{'}y$ &-8.44 & -7.93 & -7.36 & -6.78 & -6.21       \\
             & $w$          &-0.53 & -0.56 & -0.60 & -0.67 & -0.74       \\
             & $w^{'}$      &-0.40 & -0.41 & -0.43 & -0.46 & -0.49         \\
             & $u$          &0.63 & 0.66 & 0.72 & 0.79 & 0.88         \\
             & $u^{'}$      &0.47 & 0.485 & 0.51 & 0.54 & 0.58         \\
             & $x$          &-0.20& -0.21 & -0.22 & -0.25  & -0.27            \\
             & $x^{'}$      &-0.15 & -0.15 & -0.16 & -0.17 & -0.18       \\
\hline
$f^-$(ZBM-O) & ${\xi}^{'}v$ &-19.36 & -18.17 & -16.74 & -15.19 & -13.62      \\
             & ${\xi}^{'}y$ &-7.06 & -6.63 & -6.12 & -5.57 & -5.02        \\
             & $w$          &0.48 & 0.51 & 0.56 & 0.62 & 0.69       \\
             & $w^{'}$      &0.37 & 0.38 & 0.40 & 0.43 & 0.47         \\
             & $u$          &-0.56 & -0.60 & -0.65 & -0.72 & -0.81         \\
             & $u^{'}$      &-0.43 & -0.45 & -0.47 & -0.51 & -0.55         \\
             & $x$          &-0.18& -0.19 & -0.21 & -0.235  & -0.26            \\
             & $x^{'}$      &-0.14 & -0.14 & -0.15 & -0.17 & -0.18       \\
\hline
$f^-$(ZBM-O*) & ${\xi}^{'}v$ &-16.2 &-15.21 &-14.0 & -12.7 & -11.38    \\
             & ${\xi}^{'}y$ &-6.17 & -5.8  & -5.35 & -4.87 & -4.385       \\
             & $w$          &0.42 & 0.45   & 0.49  & 0.54  & 0.61     \\
             & $w^{'}$      &0.32 & 0.335  & 0.355 & 0.38  & 0.41      \\
             & $u$          &-0.49 & -0.52 & -0.57 & -0.63 & -0.705   \\
             & $u^{'}$      &-0.37 & -0.39 & -0.41 & -0.44 & -0.48        \\
             & $x$          &-0.16& -0.17  & -0.18 & -0.2  & -0.22       \\
             & $x^{'}$      &-0.12 & -0.12 & -0.13 & -0.14 & -0.15       \\
\hline
$f^-$(ZBM-F) & ${\xi}^{'}v$ &-25.61 & -24.05 & -22.16 & -20.12 & -18.05      \\
             & ${\xi}^{'}y$ &-8.87 & -8.33 & -7.69 & -7.00 & -6.30        \\
             & $w$          &0.60 & 0.64 & 0.695 & 0.77 & 0.86       \\
             & $w^{'}$      &0.46 & 0.475 & 0.50 & 0.54 & 0.59         \\
             & $u$          &-0.71 & -0.75 & -0.82 & -0.91 & -1.02         \\
             & $u^{'}$      &-0.54 & -0.56 & -0.60 & -0.64 & -0.69         \\
             & $x$          &-0.235& -0.25 & -0.27 & -0.30  & -0.34       \\
             & $x^{'}$      &-0.18 & -0.19 & -0.20 & -0.21 & -0.23       \\
\hline
 \end{tabular}
\end{center}
\end{table}

\vfill
\newpage

\begin{table}[h]

\caption{The amplitudes of components $(1s_{1/2}^20p_{1/2}^{-1})$, $(0d_{1/2}^20p_{1/2}^{-1})$
in the initial state $1/2_1^-$ of the $\beta^+/\beta^-$ decay of $^{17}$Ne/$^{17}$N and the dominant
components $1s_{1/2}$ and $0d_{5/2}$
in the final first excited state $1/2_1^+$ and in the g.s. of
$^{17}$F/$^{17}$O, respectively, for different hybrids of the ZBM interaction.
}
\label{tab4}
\begin{center}
\begin{tabular}{|c|c|c|c|c|}
\hline
Interaction & $(1s_{1/2}^20p_{1/2}^{-1})^{J^{\pi}=1/2_1^-}$ & $(0d_{5/2}^20p_{1/2}^{-1})^{J^{\pi}=1/2_1^-}$ 
& $(1s_{1/2})^{J^{\pi}=1/2_1^+}$ & $(0d_{5/2})^{J^{\pi}=5/2_1^+}$ \\
\hline
ZBM \protect\cite{zbm} & 0.41 & 0.758 & 0.65 & 0.69 \\
ZBM-F & 0.399 & 0.78 & 0.665 & 0.707 \\
ZBM-O & 0.311 & 0.80 & 0.62 & 0.667 \\
ZBM-O* & 0.268 & 0.81 & 0.587 & 0.637 \\ 
\hline
 \end{tabular}
\end{center}
\end{table}

\begin{figure}[t]
\centerline{\epsfig{figure=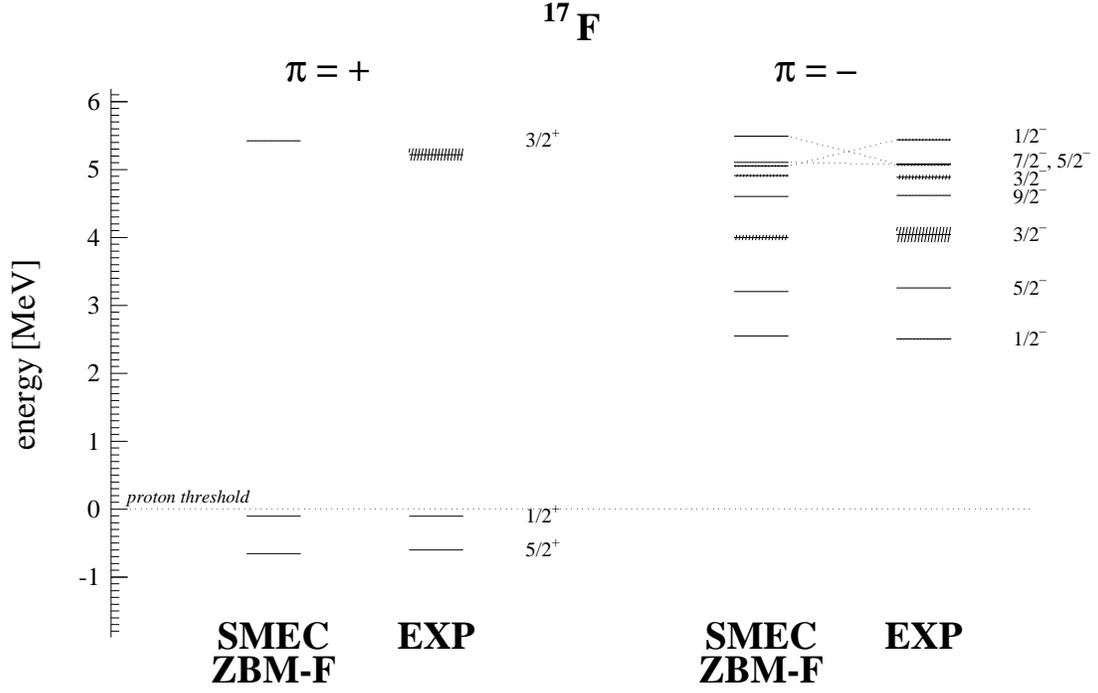,height=10cm}}
\vskip 1truecm
\caption{Comparison of experimental spectrum of $^{17}\mbox{F}$  with the
spectrum calculated using SMEC with the ZBM-F effective
interaction. Residual coupling to the continuum state
is provided by the density dependent DDSM1 interaction \protect\cite{bnop3}. 
The proton threshold
energy is adjusted to reproduce position of the $1/2_{1}^{+}$ first excited
state. The shaded regions represent the width of resonance states.}
\label{fig0d}
\end{figure}
\newpage
\begin{figure}[t]
\centerline{\epsfig{figure=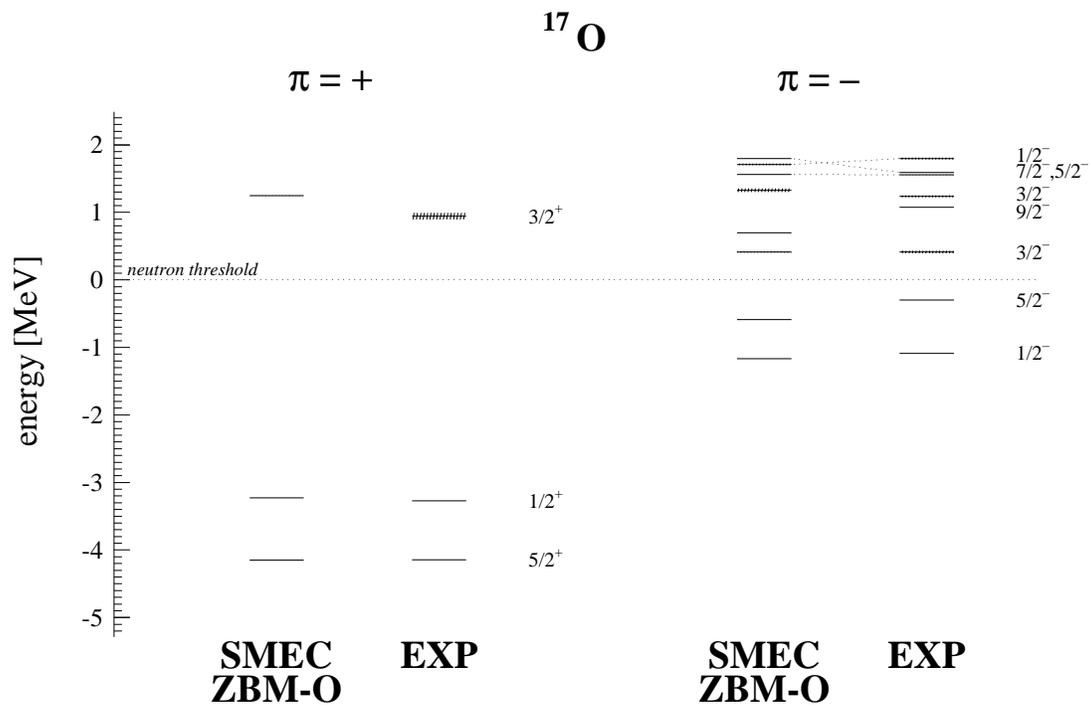,height=10cm}}
\vskip 1truecm
\caption{The same as in Fig. 1 but for $^{17}\mbox{O}$. The neutron threshold
energy is adjusted to reproduce position of the first $3/2_{1}^{-}$ excited
state.}
\label{fig0x}
\end{figure}
\newpage
\begin{figure}[t]
\centerline{\epsfig{figure=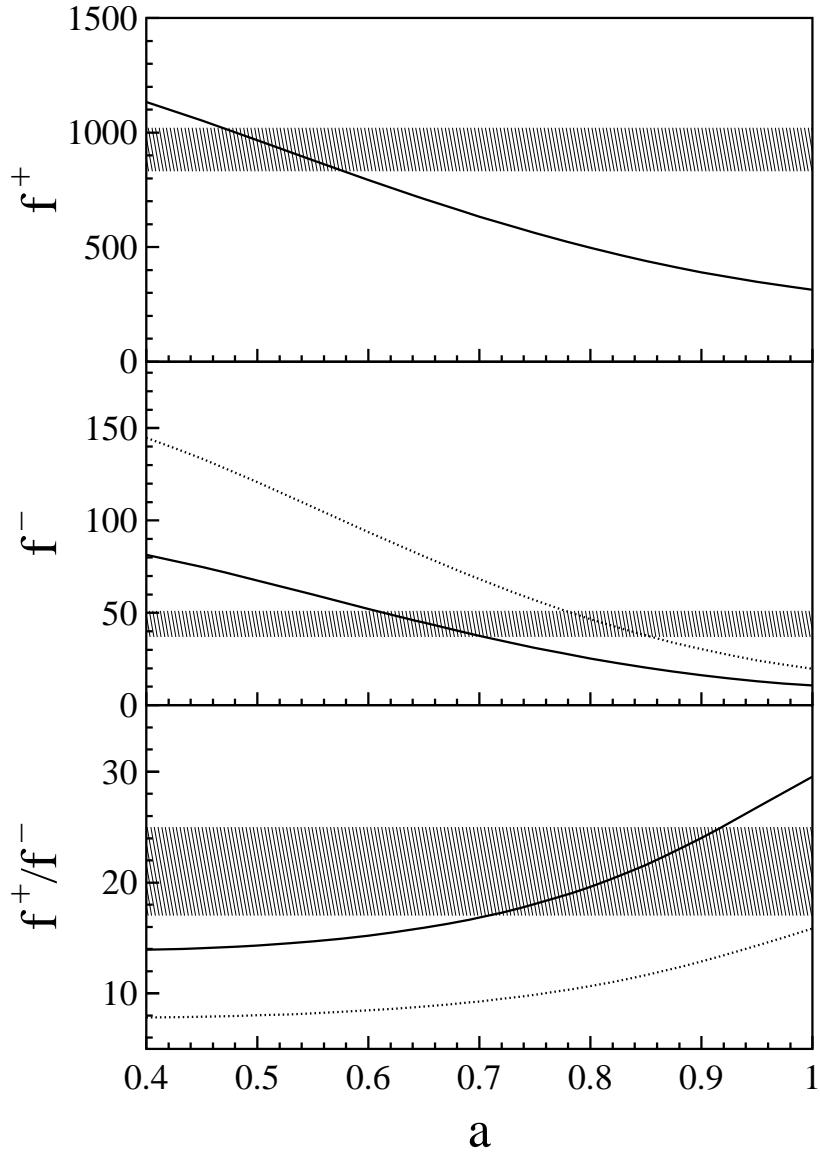,height=16cm}}
\vskip 1truecm
  \caption{%
$f^+$, $f^-$ values and the ratio $f^+/f^-$ for the first-forbidden 
$\beta-$transitions from the ground states of $^{17}$Ne and $^{17}$N,
are calculated using SMEC for different
values of the diffuseness $a$ of the initial average potential. 
Calculations are performed in the ZBM space using the DDSM1 residual 
interaction. The shaded areas show experimental uncertainty for $f^+, f^-$ and
$f^+/f^-$, respectively. The dotted line
corresponds to the mirror symmetric SMEC calculations for
$^{17}$Ne($\beta^+$)$^{17}$F and $^{17}$N($\beta^-$)$^{17}$O
using in both cases the ZBM-F effective interaction. 
}
\end{figure}
\newpage
\begin{figure}[t]
\centerline{\epsfig{figure=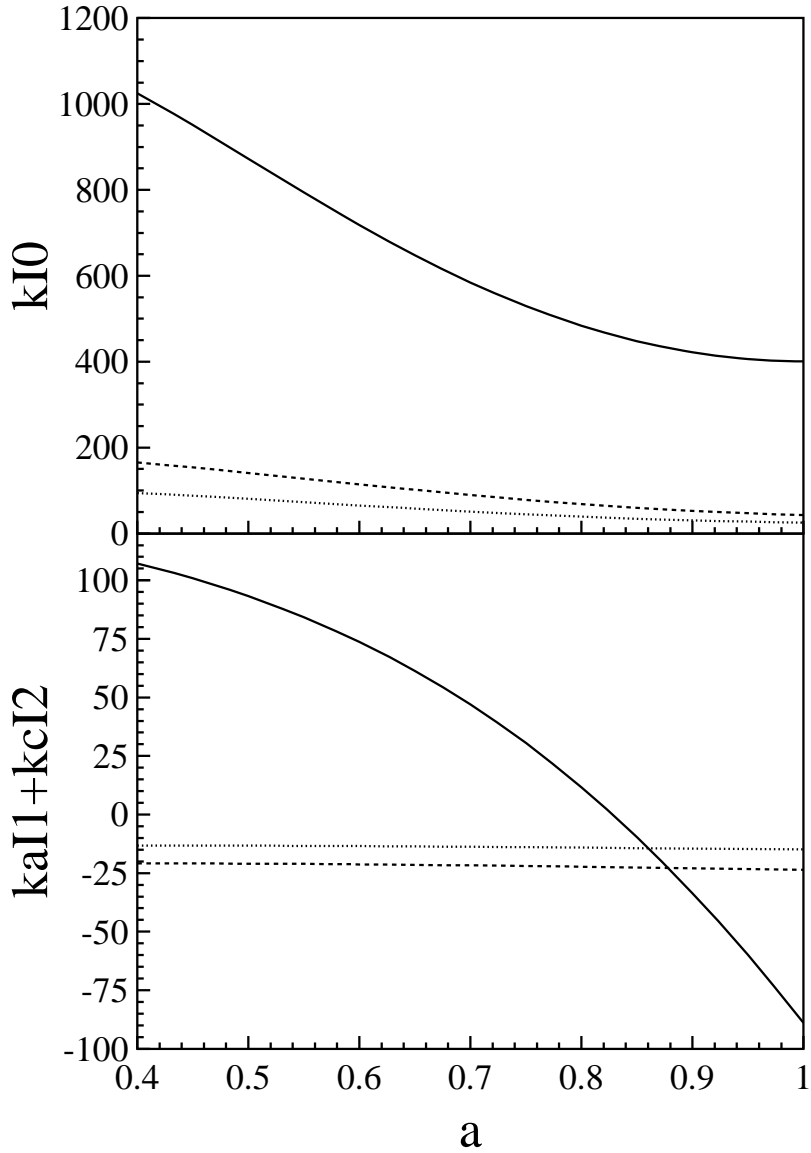,height=16cm}}
\vskip 1truecm
  \caption{%
Nuclear matrix elements $kI_0$ (upper part) and ($kaI_1+kcI_2$) (lower part) are plotted
as a function of the diffuseness of the 'first guess' initial average potential. 
Calculations are performed in the ZBM space using the DDSM1 residual 
interaction. The solid line shows the matrix elements for the transition
$^{17}$Ne(${\beta}^+$)$^{17}$F which is calculated using the ZBM-F 
interaction. The dashed and dotted lines show the matrix
elements for the transition $^{17}$N(${\beta}^-$)$^{17}$O calculated using
ZBM-F and ZBM-O interactions, respectively. 
}
\end{figure}
\newpage
\begin{figure}[t]
\centerline{\epsfig{figure=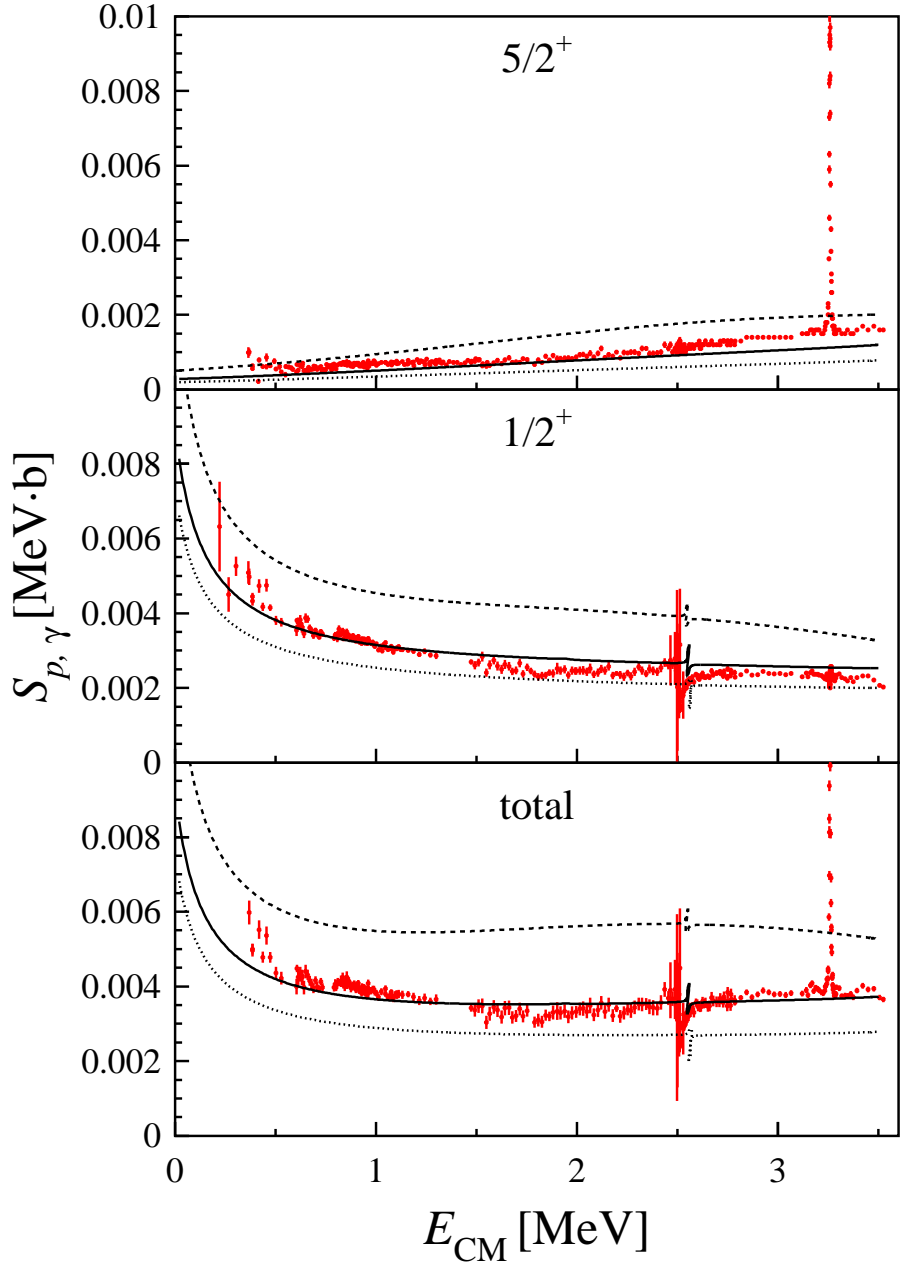,height=18cm}}
\vskip 1truecm
  \caption{%
The astrophysical $S$-factor for the reactions
$^{16}\mbox{O}(p,\gamma)^{17}\mbox{F}(J^{\pi}=5/2_{1}^{+})$ and
$^{16}\mbox{O}(p,\gamma)^{17}\mbox{F}(J^{\pi}=1/2_{1}^{+})$, is plotted
as a function of the center of mass energy $E_{CM}$ for three different values
of the diffuseness of the initial (auxiliary) potential : $a=0.4$ fm (the dotted
line), $a=0.55$ fm (the solid line) and $a=0.8$ fm (the dashed line). 
The experimental data are taken from Ref. \protect\cite{morlock}. }
\end{figure}
\newpage
\begin{figure}[t]
\centerline{\epsfig{figure=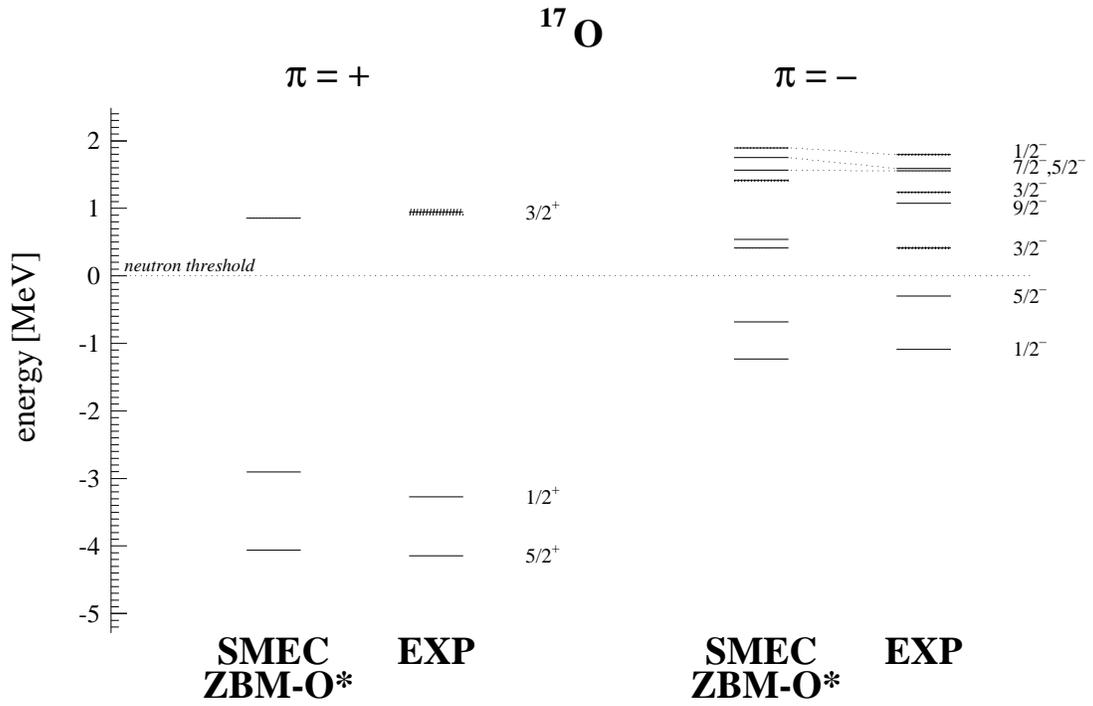,height=10cm}}
\vskip 1truecm
\caption{The same as in Fig. 2 but for ZBM-O* effective interaction
which yields the correct value for the $\beta^-$ first-forbidden decay rate 
of $^{17}$N in the ground state. The neutron threshold
energy is adjusted to reproduce position of the first $3/2_{1}^{-}$ excited
state.}
\label{fig0x}
\end{figure}
\newpage
\begin{figure}[t]
\centerline{\epsfig{figure=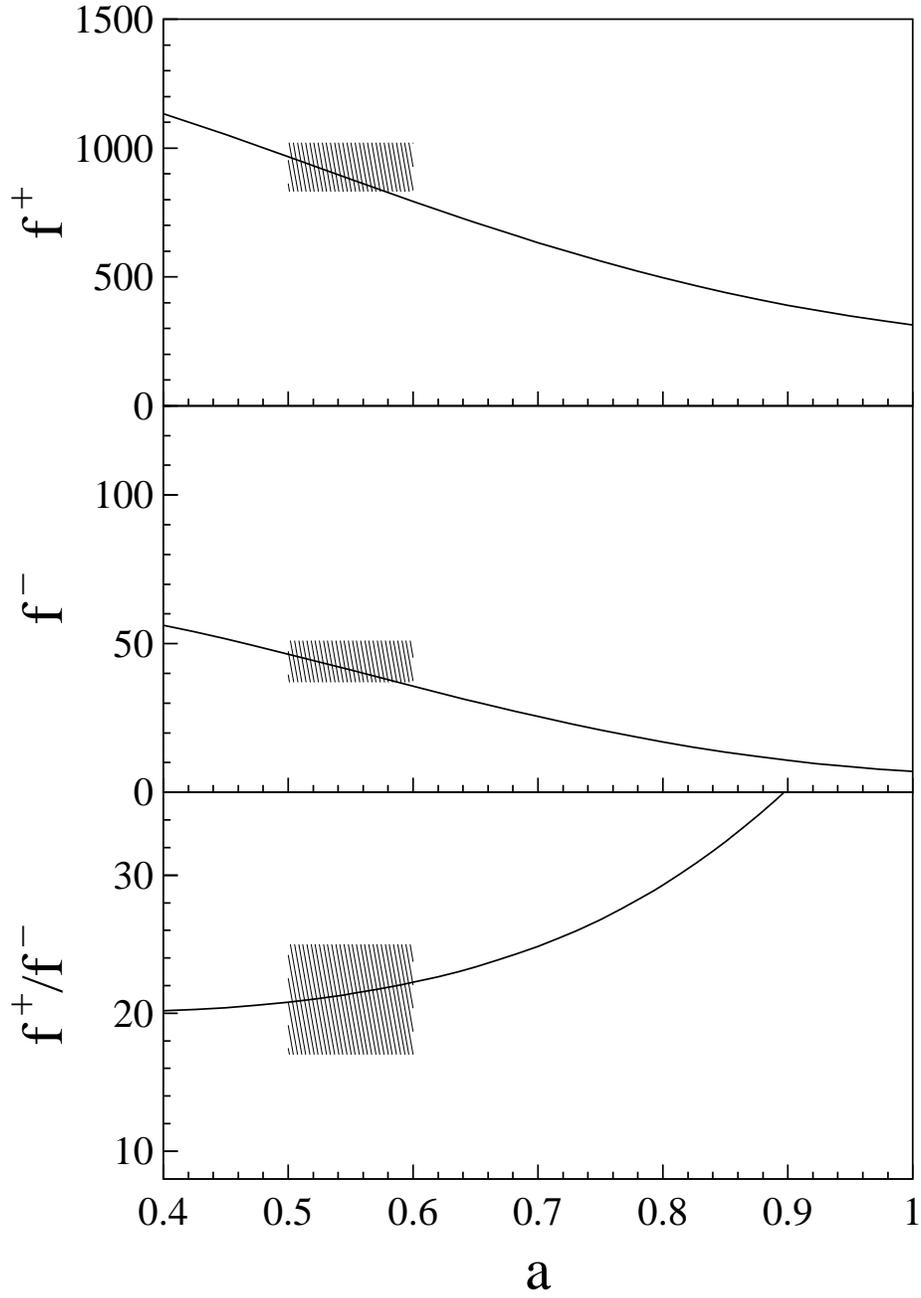,height=18cm}}
\vskip 1truecm
  \caption{%
$f^+$, $f^-$ values and the ratio $f^+/f^-$ for the first-forbidden 
transitions from the ground states of $^{17}$Ne and $^{17}$N, are calculated using SMEC for different
initial values of the diffuseness of the initial average  potential. 
ZBM-O* effective shell model interaction is used to
calculate the structure of $^{17}$O. For more details, see the description in
the text and the caption of Fig. 3.
}
\end{figure}

\end{document}